\documentclass[11pt]{article} 
 
\usepackage{amstex}
\usepackage{amscd, amssymb}
\usepackage{longtable, epsfig}
\usepackage{array}
\usepackage{oldgerm}

\begin{document}

\begin{titlepage}
\title{
\vskip -70pt
\begin{flushright}
{\normalsize \ DAMTP-2002-94}\\
\end{flushright}
\vskip 20pt \vspace{0.5cm}
{\bf The dynamics of vortices on $S^2$\\ near the Bradlow limit}
\vspace{0.3cm}}

{\makeatletter
\author{{J. M. Baptista}\thanks{e-mail address: J.M.Baptista@damtp.cam.ac.uk}
 $\ $and 
 {N. S. Manton}\thanks{e-mail address: N.S.Manton@damtp.cam.ac.uk} \vspace{0.15cm}\\
{\small {\sl Department of Applied Mathematics and Theoretical Physics}}\\
{\small {\sl University of Cambridge}} \\
{\small {\sl Wilberforce Road, Cambridge CB3 0WA, England}}
}
\makeatother}
\date{July  2002}
\maketitle
\thispagestyle{empty}
\vspace{1cm}
\vskip 20pt
{\centerline{{\large \bf{Abstract}}}}
\vspace{.35cm}
The explicit solutions of the Bogomolny equations for $N$ vortices on
a sphere of radius $R^2>N$ are not known. In particular, this has
prevented the use of the geodesic approximation to describe the low
energy vortex dynamics. In this paper we introduce an approximate
general solution of the equations, valid for $R^2 \gtrsim N$, which has
many properties of the true  solutions, including the same moduli space
$\mathbb{CP}^N$. Within the framework of the geodesic approximation, the
metric on the moduli space is then computed to be proportional to the
Fubini-Study metric, which leads to a complete description of the particle
dynamics.

\end{titlepage}

\section{Introduction}

The abelian Higgs model in the plane is one of the most studied examples
of a field theory with topological solitons. The solitons are vortices. At
critical coupling there are Bogomolny equations, and it is known that
there is a $2N$-dimensional manifold of gauge-inequivalent $N$-vortex 
solutions \cite{JT}. This is known as the $N$-vortex moduli space, and denoted
${\cal M}_N$. As a manifold, ${\cal M}_N \cong {\mathbb C}^N$.
There is a natural metric on ${\cal M}_N$, arising from the
kinetic terms in the Lagrangian of the model, and it has been proved by
Stuart \cite{Stu1} that, at least for finite time intervals, geodesic
trajectories on the moduli space give a good approximation to the true dynamics
of slowly moving vortices.

It is convenient to introduce the standard complex coordinate $z$ on the
plane. The locations of the vortices are the $N$
unordered points where the Higgs field vanishes. These points may be
regarded as the roots of a monic polynomial in $z$ (monic means that the
coefficient of $z^N$ is 1), and the natural coordinates on moduli space
are the $N$ complex coefficients of such a polynomial. If a
particular geodesic motion is known, then the time-dependence of the
polynomial is known, and hence the time-dependence of the roots can
be calculated.

Now, a general formula for the metric on moduli space has been given by
Samols \cite{Sa}, but it is not explicit, so only very special geodesics, with a
high degree of symmetry, are understood in detail for $N>2$ \cite{AB, McK}.
One vortex just moves at constant speed along a straight line.
The geodesic motion for two vortices has been calculated by Samols numerically. 
The most interesting phenomenon is that, in a head-on collision, two vortices
scatter at right angles. Recently, Manton and Speight \cite{MS} have found an
explicit metric for $N$ well separated vortices, from which the geodesic
motion could be computed.

In this paper we are interested in the opposite limit. It is possible to
consider the abelian Higgs model with fields defined on any compact
surface. We shall only consider the case of a 2-sphere with its standard
round metric, parametrized by its radius $R$. There are again Bogomolny
equations and a $2N$-dimensional moduli space of $N$-vortex solutions.
As a manifold this is $\Bbb{CP}^N$. However, there is an important geometrical
constraint, discovered by Bradlow \cite{Br}. This is that the area of the sphere
must be greater than $4\pi N$ for non-trivial solutions of the Bogomolny
equations to exist. Equivalently, $R^2 > N$. The metric on moduli space is
known to collapse to zero size as the Bradlow limit $R^2 \searrow N$ is
approached. We shall be interested here in the case where $R^2$ is slightly
greater than $N$. One should think of this as a situation where the
vortices are densely squeezed together. We shall
present an approximate general solution of the Bogomolny equations, and using
this, calculate the metric on moduli space directly from its definition.
Again the solutions involve a polynomial, and the natural coordinates
are the complex coefficients of the polynomial. We shall find that
the metric is that of Fubini-Study, with an overall scale factor
that depends on $R^2 - N$.

The geodesics on Fubini-Study are quite simple, and hence, in principle,
the motion of vortices can be calculated straightforwardly. However, this
does involve finding the roots of polynomials with time-varying
coefficients, which is not algebraically trivial for three or more
vortices. We shall present examples, mainly of two vortex motion. We
should also remark that Stuart's proof of the validity of the geodesic
approximation for vortex motion does not extend automatically to this
regime of being close to the Bradlow limit, so our results on vortex
motion remain rather formal at this stage. The particle dynamics we
obtain at the end is, nevertheless, quite ``physical''.

The paper is structured as follows. In Section 2 we introduce the
Bogomolny equations on $S^2$, which is identified with $\Bbb{CP}^1$.
As in \cite{Br, Ga}, the equations are defined on complex
line bundles over this surface. In Sections 3 and 4 we explain our
approximation for $R^2$ close to $N$, and proceed to compute the metric on the
moduli space of the approximate solutions. The geodesics of this
Fubini-Study metric are then used in Section 5 to give an explicit
description of the $N$-vortex dynamics. Some examples of motions are
presented in Section 6, and finally in Section 7 a general result
concerning the number of vortex collisions is proved.

\section{The Bogomolny equations}

According to Bradlow's generalization of the classical vortices over
$\Bbb{R}^2$ \cite{Br}, when the base manifold is the sphere $S^2\cong \Bbb{CP}^1$, the setup for the
problem is a complex line bundle $\pi:E \rightarrow S^2 $ of degree $N$ equipped with a hermitian
metric $h$. The Higgs field $\phi$ is now a section of this bundle, and the gauge potentials are 
the local 1-forms of an $h$-compatible connection $D$ on the bundle. We will take the metric on 
$S^2$ to be $g_{R}:= R^2 \times (\text{standard metric on}\, S^2)\,$, so that the volume of 
$(S^2, g_{R})$ is $4\pi R^2$. 

Letting ${\mathcal A} := \{ h\text{-compatible connections on}\ E \}\, $ and $\,\Gamma (E) :=
\{ \text{global}\  C^{\infty}\ \text{sections}$ of$\, E\}\,$, the Bogomolny equations for $(D,\phi)
\in {\mathcal A} \times  {\Gamma (E)}$ are (\cite{Br, Ga}) : 
\begin{gather}
D^{0,1}\phi = 0          \\
F + \frac{1}{2} (|\phi|^2 _h - 1 )(\text{vol}_R) = 0      
\end{gather}
where $D^{0,1}$ is the anti-holomorphic part of $D$, $\text{vol}_R \in \Omega^2 (S^2 ,
\Bbb{R})$ is the volume form of the metric $g_R$, and $-iF$ is the globally defined curvature
form of $D$, so that $F\in \Omega ^2 (S^2 , \Bbb{R})$.

We remark that the problem does not depend essentially on $(E,h)$, because all 
complex line bundles over $S^2$ of a given degree $N$ are isomorphic. In fact, for another choice
$(E' , h')$ there will
always be an isomorphism $f:E\rightarrow E' $ 
such that $f^\ast (h') =  h $ . It is then not difficult to check that $(D,\phi)$ is a
solution of $(1)$ and $(2)$ on $(E',h')$ if and only if $(f^\ast D \,,\, f^{-1}\circ \phi )$ 
is a solution on $(E,h)$, where $f^\ast D$ is the pull-back connection on $E$.  
Notice in particular that $\phi \in \Gamma (E')$ and $f^{-1}\circ\, \phi \in \Gamma(E)$ have the 
same zeros on $S^2$, and hence correspond to the same vortex configuration.\\
$\ $

We will now define the particular pair $(E,h)$ which is to be used in the remainder of this
paper.
Let $U_1 =\Bbb{CP}^1\!\setminus\!\{[0,1]\}$, $\,U_2 =\Bbb{CP}^1\!\setminus\!\{[1,0]\}$, where 
we use the standard homogeneous coordinates $[z_0,z_1]$ for points on ${\Bbb{CP}^1}$, and let 
$\varphi_i : U_i \rightarrow \Bbb{C} $ be the standard complex charts of $\Bbb{CP}^1$
with transition functions $\varphi_1 \circ \varphi_2^{-1} = \varphi_2 \circ \varphi_1^{-1} :
z \mapsto 1/z \,$.
Define $g_{ij} :U_i \cap U_j \rightarrow U(1)$ by 
\[  g_{21} \circ \varphi_2 ^{-1} (z) = (z/|z|)^N ,\ \ \ \ g_{12} =
1/g_{21}\ ,\ \ \ \  g_{11} = g_{22} = 1 \ . \] 
Since the
$g_{ij}$ satisfy the usual cocycle conditions, it is possible to construct a complex 
line bundle $\pi :E \rightarrow \Bbb{CP}^1 $ with trivializations $\psi_i : \pi^{-1}(U_i) 
\rightarrow U_i\times \Bbb{C}$ such that $\psi_i \circ \psi_j ^{-1} (p,y) = (p\,,\,g_{ij}(p)\,
y)$.
 The hermitian metric $h$ on $E$ is defined by requiring the unitarity of the trivializations 
$\psi_i\,$, that is  $|\psi_i ^{-1}(p,y)|_h ^2 = |y|^2\,$; it is well defined because $g_{ij}$ has 
values in $U(1)$.

\newpage

We should now check that deg$\,E=N$. Define the real valued 1-forms 
$A_i \in \Omega^1(U_i,\Bbb{R})$ by
\begin{gather}
  A_i = \varphi_i^\ast A \ \ \ \ \ \ \ \text{ with} \ \ \ \ \ A := 
\frac{-iN}{2(1+|z|^2)} (\bar{z} \, dz - z\, d\bar{z}) \ \ \ \in \Omega ^1(\Bbb{C,R})\ .
\end{gather}
On $U_1\cap U_2$ one has 
\[   (\varphi_2 ^{-1})^\ast(A_1 - A_2)\  =\  \left(\frac{1}{z}\right)^\ast\!\! A\, - A \ =\  i\,
\left(\frac{|z|}{z}\right)^N d\left(\frac{z}{|z|}\right)^N \, =\ i\,(\varphi_2 ^{-1})^\ast 
(g_{12}\, dg_{21})\ ,  \]
or equivalently
\[  (-iA_1)-(-iA_2)\, =\, g_{12}\, dg_{21} \ ,    \]
which shows that the local forms $-iA_1$ and $-iA_2$ define a connection $D_N$ on $E$. The 
curvature $-iF_N$ of $D_N$ is a global 2-form on $\Bbb{CP}^1$ determined by $F_N = dA_j$ on $U_j$.
In particular, one can compute that 
\[  (\varphi_1 ^{-1})^\ast F_N \  =\ dA \ =\  \frac{iN}{(1+|z|^2)^2} \, dz \wedge d\bar{z} \ =\  
 (\varphi_1 ^{-1})^\ast (\frac{1}{2} \,\text{vol}_{\sqrt{N}})\ ,    \] 
and hence 
\[   \text{deg}\,E\  :=\  \frac{i}{2\pi}\ \int_{\Bbb{CP}^1}(-iF_N) \ =\  \frac{1}{2\pi}
 \int_{\Bbb{C}} (\varphi_1^{-1})^\ast F_N \ =\  N\ .          \]

Integrating equation $(2)$ over $\Bbb{CP}^1$, and using that $\int_{\Bbb{CP}^1}F
\,=\, 2\pi\,\text{deg}\,E \, =\,2\pi N\,$, it is clear that for $R^2< N$ the Bogomolny equations
have no solution, and that for $R^2 =N$ any solution $(D,\phi)$ must satisfy $\phi = 0$ and
$F \,=\,\frac{1}{2}\,\text{vol}_{\sqrt{N}} \,=\, F_N $. Since we have already constructed a
connection $D_N$ on $E$ with curvature $-iF_N$, we have an explicit solution of the Bogomolny
equations for the case $R^2 =N$ (which is called the Bradlow limit), and it can be shown that it
is unique up to gauge transformations.\\
$\ $

For $R^2 >N$ Bradlow has shown \cite{Br}, in a more general context, that for any solution 
$(D,\phi)$ of
$(1)$ and $(2)$, the section $\phi$ has exactly $N$ zeros (which are called vortices), counting 
multiplicities. Moreover, the moduli space ${\mathcal M}_N $ of these solutions up to gauge 
tranformations, is parametrized by the positions in $\Bbb{CP}^1$ of these $N$ vortices. Since the
vortices are indistinguishable, this moduli space is identified with $(\Bbb{CP}^1)^N / 
\textfrak{S}_N$, where $\textfrak{S}_N$ is the group of permutations of $N$ elements.

Now consider the map $\Upsilon : (\Bbb{CP}^1)^N /\textfrak{S}_N \rightarrow \Bbb{CP}^N$ defined 
in homogeneous coordinates by  
\[ \Bigl[\,[u^1,v^1]\,, \ldots , [u^N,v^N]\,\Bigr] \longmapsto \Bigl[\ \ldots , \sum_{\sigma \in 
\textfrak{S}_N}v^{\sigma (1)} \cdots v^{\sigma (k)}u^{\sigma(k+1)} \cdots u^{\sigma (N)}\,,\,
\ldots  \ \Bigr]_{0\leq k\leq N}\ .   \]
With some care, one can verify that $\Upsilon$ is a bijection. In fact, its inverse may be
described in the following way . Given $[w_0, \ldots , w_N] \in \Bbb{CP}^N$, consider the non-zero
polynomial
\[  P(z)= \sum_{k=0}^{N} (-1)^k \frac{N!}{k!(N-k)!}\,w_k \, z^{N-k}\ , \]
which has degree $l\leq N$. Calling $z_1 ,\ldots, z_l$ the complex roots of $P(z)$, one has
\[  \Upsilon ^{-1}\Bigl(\,[w_0 ,\ldots , w_N]\,\Bigr)\  =\  \Bigl[\,[1,z_1],\ldots ,[1,z_l]\, ,
\,[0,1], \ldots ,[0,1]\, \Bigr] \ .    \]
Using this bijection, ${\mathcal M}_N$ can also be identified with $\Bbb{CP}^N$.

\section{Vortices near the Bradlow limit}

Although we have an accurate description of the moduli space ${\mathcal M}_N$, the explicit form 
of the solutions $(D,\phi)$ of $(1)$ and $(2)$ is not known. In particular, this has prevented
any successful attempt to describe the dynamics of the vortices by means of the well-known 
geodesic approximation.
The purpose of this paper is to show that by replacing the exact Bogomolny equations by
two other conditions, which should be a good approximation near the Bradlow limit
$R^2 \searrow N$, one can obtain the solutions explicitly; they also have $\Bbb{CP}^N$
as their moduli space, and furthermore the dynamics of these ``pseudo-vortices'' is completely 
computable in the framework of the geodesic approximation.\\
$\ $

Since for $R^2=N$ the pair $(D_N,0)\in {\mathcal A}\times \Gamma (E)$ is an exact solution of
$(1)$ and $(2)$, we may expect that for $R^2$ close to $N$ the solutions $(D,\phi)$ will
have $D\approx D_N$ (after a gauge transformation if necessary). We therefore impose $D=D_N$ and 
look for $\phi \in \Gamma (E)$ such that :
\begin{align}
&\bullet  \text{$(D_N, \phi)$ is a solution of $(1)$, i.e.} \ \ \ D_N ^{0,1} \phi = 0\ ;    \\ 
&\bullet  \text{$(D_N, \phi)$ satisfies $(2)$ ``on average'', i.e.}  
 \ \ \int_{\Bbb{CP}^1} \left( F_N + \frac{1}{2} (|\phi|_h ^2 -1)\,\text{vol}_R\right)\ = 0\ ; 
\end{align}
(\,We note in passing that eq.(4) is analogous to the
  equation for electron wavefunctions of the first Landau level in the
  uniform  background magnetic field $F_N\,$; eq.(5) is then a
  wavefunction normalization condition.\,)
 
  We will now find the explicit solutions of $(4)$ and $(5)$, and then describe their moduli
space.
Using the local trivialization $\psi_1$ of $E$ and the chart $\varphi_1$ of $\Bbb{CP}^1$
defined before, the equation $\ D_N ^{0,1} \phi = 0\ $ over the domain $U_1$ is the same as  
$\ (\bar{\partial} -iA^{0,1})\phi_1 = 0 \ $, or explicitly, using $(3)$:
\begin{gather}
\frac{\partial \phi_1}{\partial \bar{z}}\ =\ \frac{-N\,z}{2(1+|z|^2)}\,\phi_1 \ ,
\end{gather}     
where $\phi_1 \in C^\infty (\Bbb{C})$ is the representative of $\phi$ with respect to $\psi_1\,$.

  Equation $(6)$ has the general solution $\,\phi_1 = f(z)(1+|z|^2)^{-N/2}\,$, with 
$f$ holomorphic on $\Bbb{C}$. The section $\phi$ of $E$ determined by $\phi_1$,
which is only defined over $U_1\,$, has a representative $\phi_2$ with respect to $\psi_2$ given
by  $\,\phi_2 (z) = g_{12}(z) \phi_1 (\frac{1}{z}) \,$, which is smooth on $\Bbb{C}\!\setminus\! 
\{0\}$. But 
since we are looking for global solutions of $(4)$, $\phi$ must be extensible to all of
$\Bbb{CP}^1$, and this will happen iff $\phi_2(z)$ is smoothly extensible to $\Bbb{C}$.
Writing $f$ as a Taylor series, it is then not difficult to check that this requires that
the coefficient of $z^n$ vanishes for all $n>N$. Thus any solution 
$\phi$ of $(4)$ must have a representative $\phi_1$ over $U_1$ of the form 
\begin{gather}
 \phi_1 (z) = \frac{a_0 z^N + \cdots + a_N}{(1+|z|^2)^{N/2}}\ \ ,
\end{gather}   
and conversely any $\phi_1$ of this form determines a global section $\phi$ of $E$ which is
a solution of $(4)$ over $U_1$, and by continuity over all of $\Bbb{CP}^1$.

If $\phi$ is represented by $\phi_1$ as in $(7)$, then the representative $\phi_2$ will
be 
 \[ \phi_2 (z) = \frac{a_0 + \cdots + a_N z^N}{(1+|z|^2)^{N/2}} \] 
and hence, as for $(1)$ and $(2)$, any solution $\phi$ of $(4)$ has exactly $N$ zeros over
$\Bbb{CP}^1$, counting multiplicities.\\ 
$\ $

We now turn to condition $(5)$. Using that $-iF_N$ is the curvature form of a degree $N$
bundle, $(5)$ is equivalent to 
\[ 4\pi (R^2 -N) \,=\, \int_{\Bbb{CP}^1} |\phi|_h ^2 \,\text{vol}_R \,=\, \int_\Bbb{C} 
|\phi_1|^2 \frac{2iR^2}{(1+|z|^2)^2} \ dz\wedge d\bar{z} \,=\, 4\pi R^2 \sum_{k=0} ^{N}
\frac{k!(N-k)!}{(N+1)!}\,|a_k|^2 \ ,   \]
where the last integral is calculated in the appendix for $\phi_1$ of the form $(7)$. We 
can therefore conclude that $\phi_1$ represents a solution $\phi$ of $(4)$ and $(5)$ iff 
$\phi_1$ is of the form $(7)$ and satisfies the normalization condition
\[ \sum_{k=0} ^{N} \frac{k!(N-k)!}{(N+1)!}\, |a_k|^2\, =\, 1-\frac{N}{R^2} \ \ .  \]
  Calling ${\mathcal D} \subset {\mathcal A} \times \Gamma (E)$ the subspace of solutions 
of $(4)$ and $(5)$, we thus get a bijection $\alpha : {\mathcal D} \rightarrow S^{2N+1} 
\subset \Bbb{C}^{N+1}$ that maps each $\phi \in {\mathcal D}$, represented by a $\phi_1$ like
in $(7)$, to the point  
\begin{gather}
\Big(1-\frac{N}{R^2}\Big)^{-1/2} \Big( \,\ldots\, ,\ \Big(\frac{k!(N-k)!}{(N+1)!}\Big)^{1/2}
a_{k}\, ,\, \ldots \Big)_{0\leq k \leq N}
 \ \ \ .  
\end{gather}

  The following step is to determine when two solutions in ${\mathcal D}$ are gauge 
equivalent. Let therefore $(D_N, \phi)$ and $(D_N, \tilde{\phi})$ be a
pair of solutions, an suppose $g: \Bbb{CP}^1 \rightarrow
U(1)$ is a gauge transformation on $E$ that takes one into the
other. Using the usual transformation rule for connection forms under
$g$, and the key fact that the connection is fixed, it is readily
shown that $g$ must be constant. So 
$\tilde{\phi} = e^{i\beta}\phi$ for some $\beta \in \Bbb{R}$. Since the converse is clear, we
conclude that $(D_N , \phi)$ and $(D_N, \tilde{\phi})$ are gauge equivalent iff 
\[  \tilde{\phi}\,=\, e^{i\beta}\phi\ \ \Leftrightarrow\ \ \tilde{\phi_1} \,=\, e^{i\beta}\phi_1\ 
  \ \Leftrightarrow\ \ \tilde{c}\,=\, e^{i\beta}c 
\]
for some $\beta \in \Bbb{R}$, where $c,\tilde{c} \in S^{2N+1}$ are the images of
$(D_N , \phi)$ and $(D_N , \tilde{\phi)}$ under the bijection $\alpha$. Furthermore, notice
that this last condition is also equivalent to $\pi (\tilde{c}) = \pi(c)$ in $\Bbb{CP}^N$, 
where $\pi: S^{2N+1} \rightarrow \Bbb{CP}^N$ is the usual principal $U(1)$-bundle.

Calling ${\mathcal M}_N $ the moduli space of solutions of $(4)$ and $(5)$ up to gauge 
transformations, and $p: {\mathcal D} \rightarrow {\mathcal M}_N$ the natural projection, we
therefore have :
\newcommand{\End}{\operatorname{End}}
\begin{equation}
\begin{CD}
{\mathcal D} @>\alpha>>S^{2N+1}\\
  @VV{p}V            @VV{\pi}V\\
{\mathcal M}_N @>\tilde{\alpha}>>\Bbb{CP}^N
\end{CD}
\end{equation}
where $\tilde{\alpha}$, defined by the commutativity of the diagram, is, like $\alpha$, a
bijection. The right-hand side of this diagram is a concrete model for the space of
solutions ${\mathcal D}$ and its moduli space.

 \section{The metric on the moduli space}

Using the usual prescriptions of the geodesic approximation (first
described in \cite{Ma1}), we will now obtain the
metric $m$ on ${\mathcal M}_N$ which, within the framework of this approximation, determines 
the dynamics of the ``pseudo-vortices'' (which from now on will be just called vortices).\\
$\ $

Suppose one has a curve $\gamma$ in ${\mathcal D}$:
\[  t\ \overset{\gamma}{\longmapsto}\ (D_N ,\phi (t)) \in {\mathcal D}\ \overset{\alpha}
{\longmapsto}\ (w_0(t), \ldots , w_N(t)) \in S^{2N+1}\ . \]
A natural hermitian metric on ${\mathcal D}$ is defined by
\begin{gather}
 \langle \frac{d\gamma}{dt} , \frac{d\gamma}{dt} \rangle_{\gamma (t)} := \frac{1}{2} 
 \int_{\Bbb{CP}^1} h\big(\dot{\phi}(t), \dot{\phi}(t)\big)\
 \text{vol}_R\ \,, 
\end{gather}
where the dot stands for the time derivative. Notice that in this case, as opposed to
what happens in \cite{Sa}, the gauge potentials do not contribute to the metric, since the 
connection in our space ${\mathcal D}$ is fixed, and thus time independent.

  Writing 
\[  \phi_1 (t) = \frac{a_0(t)\,z^N+\cdots+a_N(t)}{(1+|z|^2)^{N/2}}
\]
for the usual representative of $\phi (t)$, using the unitarity of $\psi_1$, and noting
that \\ $\dot{\phi}_1 = (1+|z|^2)^{-N/2} (\dot{a}_0 z^N+ \cdots +\dot{a}_N)\,$, one has that 
\begin{align*}
 \langle \frac{d\gamma}{dt} , \frac{d\gamma}{dt} \rangle_{\gamma (t)}\, & =\, \frac{1}{2} 
\int_{\Bbb{C}} |\dot{\phi}_1|^2 \,
 \frac{2iR^2}{(1+|z|^2)^2} \, dz\wedge d\bar{z} \,=\, 2\pi R^2 \sum_{k=0} ^N 
\frac{k!(N-k)!}{(N+1)!}\, \dot{a}_k \,\dot{\bar{a}}_k \, =  \\ 
& =\, 2\pi (R^2-N) \sum_{k=0} ^N \dot{w}_k\, \dot{\bar{w}}_k \ \,,
\end{align*}
again using the integral calculated in the appendix. We conclude that the hermitian $L^2$
metric on ${\mathcal D}$ corresponds via the map $\alpha$ to the restriction to $S^{2N+1}$ of the 
canonical hermitian metric on $\Bbb{C}^{N+1}$, up to the constant factor $2\pi (R^2 -N)$.
This metric will also be called $\langle \cdot ,\cdot \rangle$. \\
$\ $

  According to the usual procedure, the metric $m$ on ${\mathcal M}_N$ is induced from 
$\langle \cdot , \cdot \rangle$ on ${\mathcal D}$ in the following way.
Given $q \in {\mathcal D}$ and a tangent vector $\frac{d\gamma}{dt} \in T_q{\mathcal D}\,$, 
let $(\frac{d\gamma}{dt})_\perp$ be its
component perpendicular to the subspace of $T_q{\mathcal D}$ formed by the vectors 
tangent to curves on ${\mathcal D}$ which are pure gauge transformations, that is perpendicular 
to ker$(p_\ast)_q$. Then 
\[  (p^\ast m)_{q} (\frac{d\gamma}{dt},\frac{d\gamma}{dt})\, := \,\langle 
(\frac{d\gamma}{dt})_\perp , (\frac{d\gamma}{dt})_\perp \rangle_q \ \ . \] 
  We will now compute the metric on $\Bbb{CP}^N$ corresponding to $m$ by the 
identification $\tilde{\alpha}$. It will also be called  $m$.

  Using the diagram $(9)$, the subspace ker$(p_\ast)_q \subset T_q {\mathcal D}$ corresponds
to the subspace ker$(\pi _\ast)_{\alpha (q)} \subset T_{\alpha(q)}S^{2N+1}$. Given 
$w \in S^{2N+1} \subset \Bbb{C}^{N+1}\,$, we have that $\,$ker$(\pi_\ast)_w$ is the 
one-dimensional real subspace generated by 
the vector $\frac{d}{dt} (e^{it}w) (0) = iw $. Therefore, given a tangent vector  
$\frac{d\gamma}{dt} = (\dot{\gamma}_0 , \ldots, \dot{\gamma}_N )\, \in\,
T_w S^{2N+1} \subset T_w {\mathbb C}^{N+1}$, we have 
\[ 
(\frac{d\gamma}{dt})_\perp = \frac{d\gamma}{dt} - 
\frac{\langle \frac{d\gamma}{dt}, w \rangle} {\langle w,w \rangle} w\ \ ,  
\]
so
\[  
\langle (\frac{d\gamma}{dt})_\perp , (\frac{d\gamma}{dt})_\perp \rangle_w \,=\,
\langle \frac{d\gamma} {dt}, \frac{d\gamma}{dt} \rangle  - 
\frac{\langle \frac{d\gamma}{dt} ,w \rangle \langle w,\frac{d\gamma}{dt} \rangle }
{\langle w,w \rangle } \,=\, 2\pi (R^2 -N) \sum_{j,k=0} ^N (\delta _{jk} - \bar{w}_j w_k)
\,\dot{\gamma}_j \,\dot{\bar{\gamma}}_k \ \ , 
\]
where the computation in the last step uses that $ \langle w,w \rangle = 2\pi(R^2 -N)$, 
since $w \in S^{2N+1}$. Thus $\pi^\ast m$ is the restriction to $S^{2N+1}$ of the 
$2$-tensor in $\Bbb{C}^{N+1}$
\[  2\pi(R^2- N) \sum_{j,k=0}^N (\delta_{jk} - \bar{w}_j w_k)\, dw_j\otimes d\bar{w}_k \ \ 
\ .\]
Now consider the K\"ahler form $\mu$ associated to $m$. It is
defined, as usual, using the imaginary part of $m$: 
$\ \mu = -\text{Im}\,m\ \in\,\Omega^2 (\Bbb{CP}^N ,\Bbb{R})$. We have 
\begin{align*}
\pi^\ast \mu & = \pi^\ast (-\text{Im}\,m)\,=\, -\text{Im}(\pi^\ast m)\,=\, 2\pi (R^2 -N) 
\frac{i}{2} \sum_{j,k=0}^N (\delta_{jk} - \bar{w}_j w_k)\, dw_j\wedge d\bar{w}_k\ \ 
\arrowvert_{S^{2N+1}} \\ & =    2\pi (R^2 -N) \frac{i}{2} \sum_{j,k=0}^N \left(\frac{\delta_{jk}}
{|w_0|^2+\cdots +|w_N|^2} - \frac{\bar{w}_j w_k}{(|w_0|^2+\cdots+|w_N|^2)^2} \right)\,
 dw_j\wedge d\bar{w}_k\ \ \arrowvert_{S^{2N+1}} \\ & =  2\pi (R^2 -N) \frac{i}{2}\, \partial 
\bar{\partial} \log (|w_0|^2+\cdots+|w_N|^2)\ \ \arrowvert_{S^{2N+1}} \,=\,
2\pi (R^2-N)\, \pi^\ast \mu_{\rm{FS}} 
\end{align*}   
where $\mu_{\rm{FS}}$ is the Fubini-Study form on $\Bbb{CP}^N$, and the last equality is a
well-known result \cite[p. 160]{KN}. Since $\pi$ and $(\pi_\ast)_w$ are both surjective,
\begin{equation*}
  \pi^\ast \mu = \pi^\ast (2\pi (R^2 -N)\mu_{\rm{FS}})\ \ \ \ {\rm implies}\ \ \ \ 
\mu = 2\pi (R^2 -N)\mu_{\rm{FS}}\ \ .
\end{equation*}
Therefore $m=2\pi (R^2 -N)m_{\rm{FS}}\,$, where $m_{\rm{FS}}$ is the
Fubini-Study metric on $\Bbb{CP}^N$, because a hermitian metric is uniquely determined by 
its K\"ahler form.

\section{Vortex dynamics}

  Having determined the metric $m$ on the moduli space ${\mathcal M}_N \cong \Bbb{CP}^N$,
we will now proceed to explicitly describe its geodesics, which 
provide an approximate description of the low-energy particle dynamics.
 Note that $m \propto m_{\rm{FS}}$ implies that
the geodesics of $m$ are exactly the Fubini-Study geodesics. These are well-known 
\cite[p. 277]{KN} but nevertheless we will rederive them here again.

  Let $\pi : \Bbb{C}^{N+1}\! \setminus \!\!\{0\} \longrightarrow \Bbb{CP}^N$ be the natural 
projection and $\chi_{0} : U_{0} \longrightarrow \Bbb{C}^N$ one of the standard 
charts of $\Bbb{CP}^N$, where $U_{0} = \{ [w^0, \ldots, w^{N}] \in \Bbb{CP}^N :
w^{0} \neq 0\}$. Calling $(c^1,\ldots ,c^N)$ the coordinate functions of this chart,
then by definition of the Fubini-Study metric we have on $U_0\,$:
\begin{align*}
\mu_{\rm{FS}} \,& =\, \frac{i}{2} \partial \bar{\partial} \log (1+
|c^1|^2+ \cdots + |c^N|^2) \,=\, \frac{i}{2} h_{j\bar{k}} \,dc^j \wedge d\bar{c}^k\ \ \ 
\ \ \ \ \ \ \ \text{and} \\ 
 m_{\rm{FS}}\, & =\,h_{j\bar{k}} \,dc^j \otimes d\bar{c}^k\ \ ,
\end{align*}
with
\begin{gather}
h_{j\bar{k}}\, =\, \frac{\delta_{jk}}{1+|c|^2} - \frac{c^k \bar{c}^j}{(1+|c|^2)^2}\ \ . 
\end{gather}  
For a general K\"ahler metric the geodesic equations have the simplified form 
\cite[p. 4]{Ti}:
\[ \ddot{c}^k \,=\, - \frac{\partial h_{j\bar{s}}}{\partial c^l}\, h^{k\bar{s}}\,
\dot{c}^l\, \dot{c}^j \ \ ,\ \ \ \ \ \ \ \ \text{ where}\ \ \ h^{k\bar{s}}:= 
(\text{$sk$ entry of $[h_{i\bar{j}}]^{-1}$})\ \ . 
\]
  In our case $h^{k\bar{s}} = (1+|c|^2)(\delta^{sk} + c^k \bar{c}^s)$, and so the 
geodesic equations for $(\Bbb{CP}^N, m_{\rm{FS}})$ in the chart $\chi_{0}$ are 
\begin{gather}
 \ddot{c}\, = \,\frac{2<\dot{c}\,,c>}{1+<c\,,c>}\, \dot{c} \ \ ,
\end{gather}
 where $c(t)$ is a curve in $\Bbb{C}^N$ and $<\cdot , \cdot>$
is the canonical hermitian product.

  Now consider curves in $S^{2N+1} \subset \Bbb{C}^{N+1}$ of the form:
\begin{gather}
 \gamma (t) \,=\, \sin (\omega t)\,y + \cos (\omega t)\, x  \ \ ,\ \ \ \ \ \ \ \ \ 
\ \ \ t \in\Bbb{R}   
\end{gather}
where $x,y \in \Bbb{C}^{N+1}$ are orthonormal with respect to the canonical hermitian 
product of $\Bbb{C}^{N+1}$. 
If $x^{0} \neq 0$, then $\pi \circ \gamma (t) \notin U_{0}$ only for a discrete
set $D$ of non-zero values of $t$, and a short computation shows that, in $\Bbb{R}\!
\setminus \!D$, 
\begin{gather}
 c(t)\,:= \,\chi_{0} \circ \pi \circ \gamma (t) \,=\, c(0) + \dot{c}(0) \frac{x^{0}
 \sin (\omega t)}{\omega (y^{0}\sin (\omega t) + x^{0}\cos (\omega t))}  
\end{gather}
where 
\begin{gather}
c^k (0) \,=\, \frac{x^k}{x^{0}} \ \ \ \ \ \text{and} \ \ \ \ \ \ \ \dot{c}^k (0) \,=\,
\frac{\omega (y^k x^{0} - x^k y^{0})}{(x^{0})^2}  \ \ ,\ \ \ \ \ k=1,\ldots,N\ \ . 
\end{gather}

One can verify directly that this $c(t)$ satisfies $(12)$, and therefore $\pi \circ
\gamma (t)$ is a geodesic for $t$ in $\Bbb{R}\!\setminus \!D$, and by continuity for all
$t$. If $x^{0} = 0\,$, a similar computation in one of the other standard charts of $\Bbb{CP}^N$
would establish that, also in this case, $\pi \circ \gamma (t)$ is a geodesic.

  On the other hand, it is not difficult to verify that every geodesic of $(\Bbb{CP}^N 
, m_{\rm{FS}})$ can be written as $\pi \circ \gamma$, where $\gamma$ has the form $(13)$. 
Although one could give a general, chart-independent argument for this, for later 
convenience we will proceed unnaturally. Namely, using $(15)$, one may simply check that 
given any $c(0) \in \Bbb{C}^N$ and $\dot{c}(0) \in T_{c(0)}\Bbb{C}^N \cong \Bbb{C}^N$,
the geodesic $\pi \circ \gamma (t)$ with 
\begin{equation}
\begin{split}
x \,& =\, (1+|c(0)|^2)^{-1/2} \ \big(\,1\,,\,c(0)\,\big)      \\
y \,& =\, \omega^{-1} (1+|c(0)|^2)^{-1/2} \ \bigg(\,-\frac{<\dot{c}(0), c(0)>}{1+|c(0)|^2}\ 
,\ \dot{c}(0) - \frac{<\dot{c}(0), c(0)>}{1+|c(0)|^2}\, c(0)\ \bigg)      \\
   \omega \,& =\, (1+|c(0)|^2)^{-1/2} \ \bigg(|\dot{c}(0)|^2 - \frac{|<\dot{c}(0), c(0)>|^2}
              {1+|c(0)|^2} \bigg)^{1/2}             
\end{split}
\end{equation}
has initial position and velocity $\chi_{0} ^{-1} (c(0))$ and $(\chi_{0} ^{-1})_\ast
(\dot{c}(0))$, respectively. This shows that every geodesic starting in $U_{0}$ is of 
the form $\pi \circ \gamma (t)$. Similar calculations in the other standard charts 
would extend the result to all of $\Bbb{CP}^N$.

  We now note two general properties of the geodesics $\pi \circ \gamma (t)$. Firstly,
using $(11)$ and $(15)$, one can compute that the velocity of the geodesic,
which is a constant of motion, is $|\omega |$. Secondly, notice that all the
geodesics of $(\Bbb{CP}^N , m_{FS})$ are periodic. It is not difficult to show that 
for $\omega \neq 0$ the period is $\pi /|\omega |$.\\
$\ $ 

  We will now use our knowledge of the geodesics on the moduli space $(\Bbb{CP}^N , m)$ 
to give an explicit description of the vortex dynamics.

  Recall from Section 3 that the solutions $(D_N , \phi) \in {\mathcal D}$ with the
vortices (zeros of $\phi$) in positions $\varphi_1 ^{-1} (z_1), \ldots ,
\varphi_1 ^{-1} (z_N)$ in $\Bbb{CP}^1$, are represented by a function $\phi_1$ of the 
form $(7)$, where we now have 
\[  a_0 z^N + \cdots + a_N \ \propto \ (z-z_1)\cdots(z-z_N)\ \ ,\ \]
and therefore 
\[  a_k \,=\, A\, (-1)^{k}\, S_{k}(z_1, \ldots , z_N)\ \ ,\ \ \ \ \  \ k=0,\ldots,N \]
where the $S_j$ are the usual elementary symmetric polynomials, and $A$ is a normalization
factor which is non-zero for $R^2 > N$. Thus such solutions $(D_N , \phi) \in
{\mathcal D}$ correspond by $\pi \circ \alpha$ to (see$\,(8)\,$)
\[ \bigg[\ \ldots, (-1)^{k}\, \Big(\frac{k!(N-k)!}{(N+1)!} \Big)^{1/2} S_{k}(z_1, 
\ldots, z_N )\,, \ldots \ \bigg]_{0\leq k\leq N} \ \in\ \Bbb{CP}^N \cong {\mathcal M}_N 
 \]
and by $\ \chi_{0} \circ \pi \circ \alpha\ $ to $\ c=(c^1, \ldots, c^N) \,\in
\Bbb{C}^N$, where 
\begin{gather}
 c^k = (-1)^{k}\, \binom{N}{k}^{-1/2} S_{k}(z_1,\ldots,z_N) \ \ .
\end{gather}  

  Inverting these relations, we can obtain the positions of the $N$ vortices as a function 
of the coordinates $c^k$ of a given point in the moduli space $\Bbb{CP}^N$. In
particular, to the geodesics $c(t)$ of the form $(14)$ corresponds a motion of the vortices
determined by: 
\begin{gather}  
w^N + \sum_{k=1}^N \binom{N}{k}^{1/2} c^k (t)\, w^{N-k} \,=\, (w-z_1(t))\cdots
(w-z_N(t))         
\end{gather}
where the $z_i(t)$ are the coordinates of the vortices in the chart $(\varphi_1\,, U_1)$ 
of $\Bbb{CP}^1$. Thus, since we know all the geodesics of $(\Bbb{CP}^N , m)$,
we can determine all the possible $N$-vortex motions by finding the roots of polynomials of
degree $N$ --- either analytically for $N\leq 4$ or numerically for $N>4$. \\
 $\ $ 

  Now suppose we are given initial positions $z_i(0)$ and initial velocities
$\dot{z}_i (0)$ for the vortices, 
where we assume that the $z_i(0)$ are all different. Through $(17)$ and its derivative 
we can get the corresponding values $c(0),\dot{c}(0) \in \Bbb{C}^N$, then use $(16)$ to 
determine which geodesic $c(t)$ corresponds to this initial data, and finally solve 
$(18)$ to get the motions $z_i(t)$. This general procedure has been used to obtain the 
various special vortex motions shown below.

  We remark that, because $(17)$ is a local diffeomorphism only in the region where the 
vortices do not coincide, only in this region can we guarantee that the final result 
$z_i(t)$ has indeed the prescribed initial velocities. This is why we take the $z_i(0)$ all 
different. If the $z_i(0)$ are not all different there are some values
of $\dot{z}_i(0)$ that do not correspond to any $z_i(t)$ coming from a geodesic motion.

\section{Examples of motions}

Using the method described in the previous Section, we now give a few 
examples of 2-and 3-vortex motions on the sphere . The trajectories are
shown in the complex plane through the use of the stereographic 
projection $\varphi_1 : S^2\! \setminus\!\{N\} \rightarrow \Bbb{C}$.
The particular initial positions and velocities used in each case are listed in Table 1.
\setlength{\extrarowheight}{2pt}
\begin{longtable}{|c||c|c|c|c|c|c|c|c|c|}  \hline
 $\ $          & {\bf 1(a)}&{\bf 1(b)}&{\bf 1(c)}&{\bf 1(d)}&{\bf 2(a)}
 &{\bf 2(b)}    &{\bf 2(c)}  &{\bf 2(d)}    &{\bf 3} \\\hline\hline
 ${\bf z_1 (0)}$     &$1+i$  &$1+i$  &$ 1$ &$-2\,i/\sqrt{3}$&$1/2$ &$1/2$     &$1/2$ &$1/2$   &$a>0 $  \\
 ${\bf \dot{z}_1(0)}$&$-1-i$ &$-1-i$ &$-3$ &$2\,i/\sqrt{3}$&$i$     &$i$     &$i$   &$i$     &$\ i$  \\\hline
 ${\bf z_2 (0)}$     &$0$    &$-1-i$ &$-1$ &$1+i/\sqrt{3}$ &$2$     &$2$     &$2$   &$2$     &$-a$  \\
 ${\bf \dot{z}_2(0)}$&$0$    &$1+i$  &$ 1$ &$-1-i/\sqrt{3}$&$0.6\,i$&$3.7\,i$&$4\,i$&$4.5\,i$&$-i$  \\\hline
 ${\bf z_3 (0)}$     & ---   & ---   & --- &$-1+i/\sqrt{3}$&---     &---     &---   &---     &---  \\
 ${\bf \dot{z}_3(0)}$& ---   & ---   & --- &$1-i/\sqrt{3}$ &---     &---     &---   &---     &---  \\\hline
\caption{{\it Initial positions and velocities}}
\end{longtable} 
To help with the interpretation of the figures, we recall that the 
stereographic projection has the property of mapping circles of
$S^2$ (not necessarily great circles) to circles and straight lines
in the plane. The inverse $\varphi_1 ^{-1}$ has the converse 
property. Also the circle of unit radius is always shown in a dashed line;
 the northern (southern) hemisphere of $S^2$
projects to the exterior (interior) of that circle.

$\ $
$\ $\\ 

\begin{minipage}[b]{.46\linewidth}
\centering\epsfig{figure=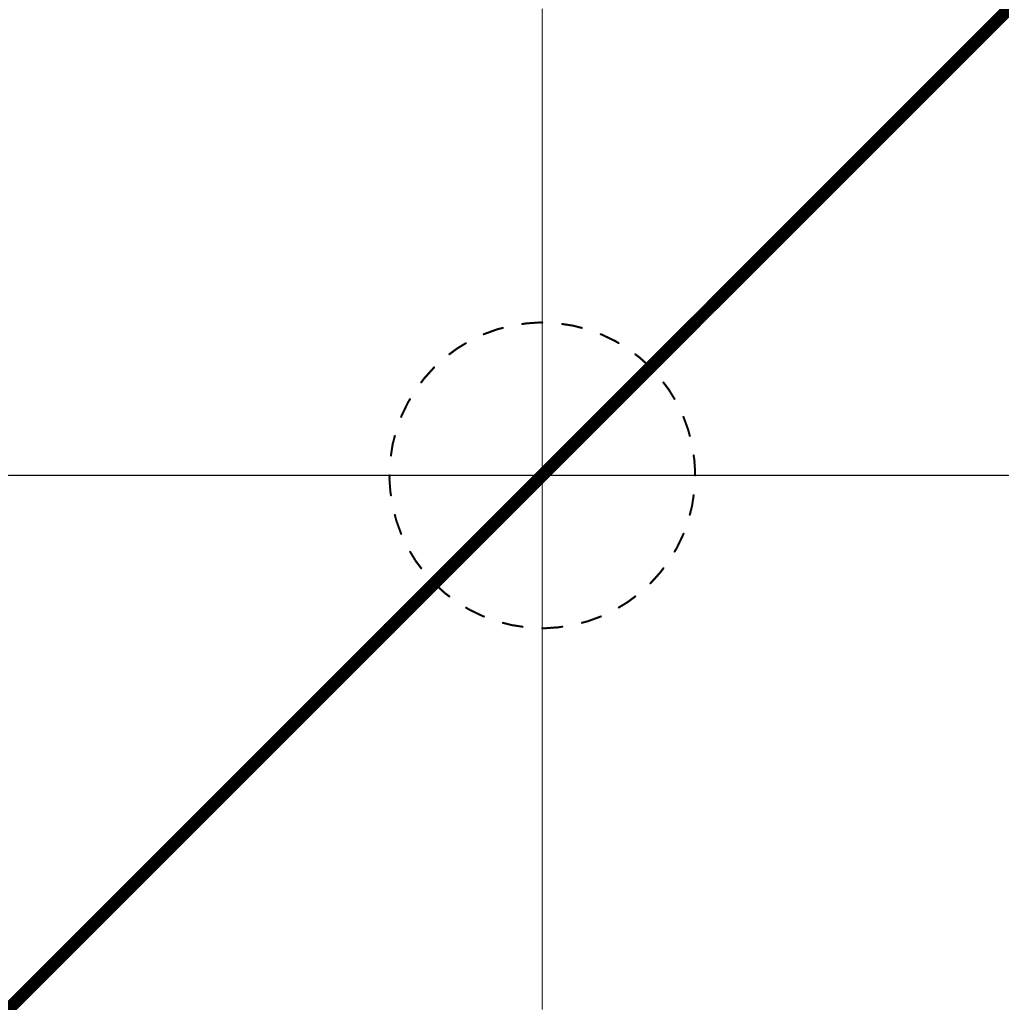 , width=3.5cm}
{\normalsize Fig. 1(a)}
\end{minipage} \hfill
\begin{minipage}[b]{.46\linewidth}
\centering\epsfig{figure=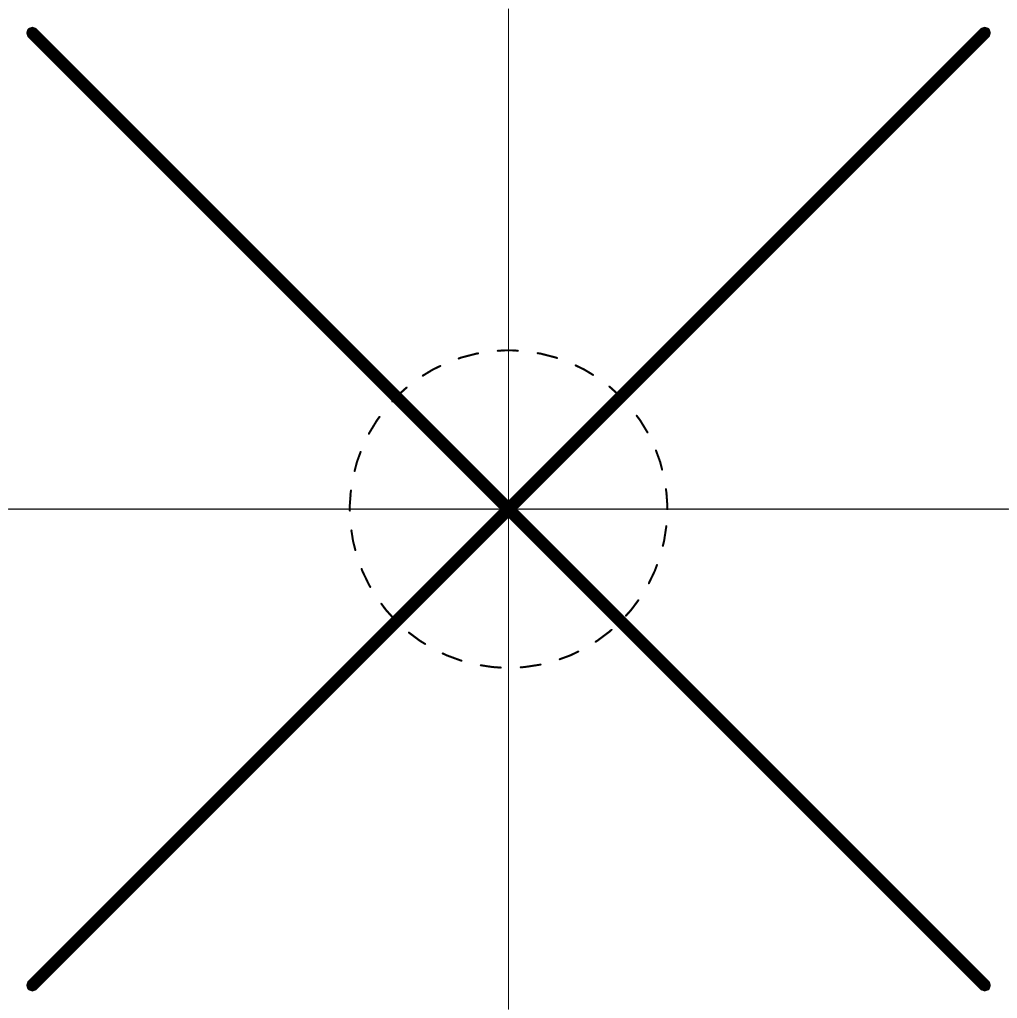 , width=3.5cm}
{\normalsize Fig. 1(b)}
\end{minipage}

$\ $\\
$\ $

\begin{minipage}[b]{.46\linewidth}
\centering\epsfig{figure=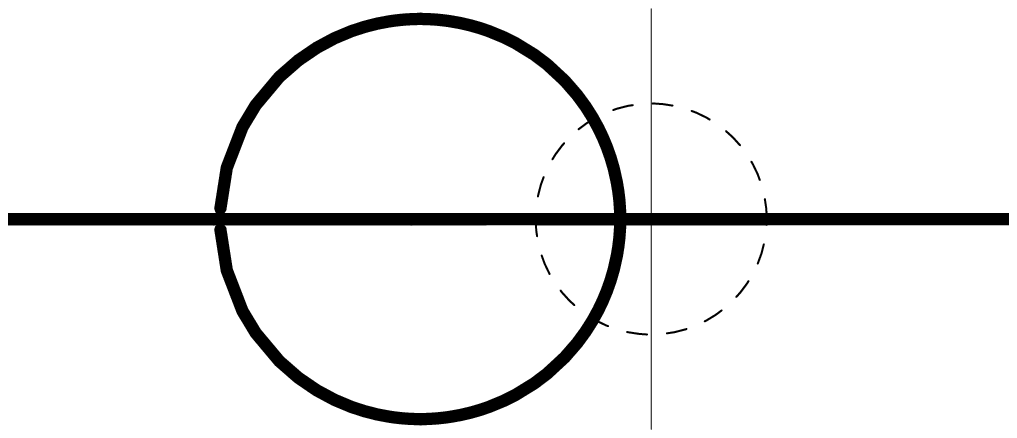 , width=3.5cm}
{\normalsize Fig. 1(c)}
\end{minipage} \hfill
\begin{minipage}[b]{.46\linewidth}
\centering\epsfig{figure=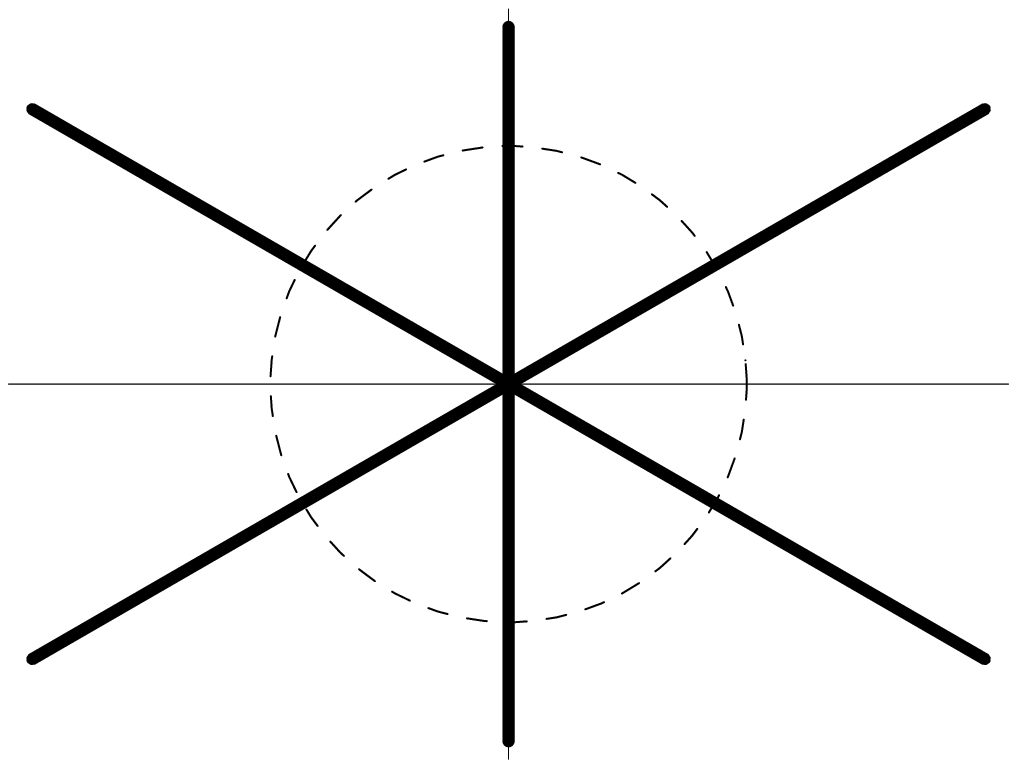 , width=3.5cm}
{\normalsize Fig. 1(d)}
\end{minipage}

$\ $

{\large \bf Figure 1 :}
\begin{description}
\item[(a)]  Two colliding vortices, one of which is at rest.\\
One of the particles describes a great circle on the sphere 
that passes through the static position of the other.

\item[(b)] Head-on collision of two vortices with the same
speed.\\ There are two collisions at antipodal points. The total 
trajectory is the union of two great circles.

\item[(c)] Head-on collision of two vortices with different 
initial speeds.\\ Again two collisions occur. The total trajectory is
the union of a great circle and another circle.

\item[(d)] Symmetrical collision of three vortices with equal 
speeds. \\ The three vortices collide twice at antipodal 
points. The total trajectory is the union of three great 
circles.

\end{description}

$\ $

\begin{minipage}[b]{.46\linewidth}
\centering\epsfig{figure=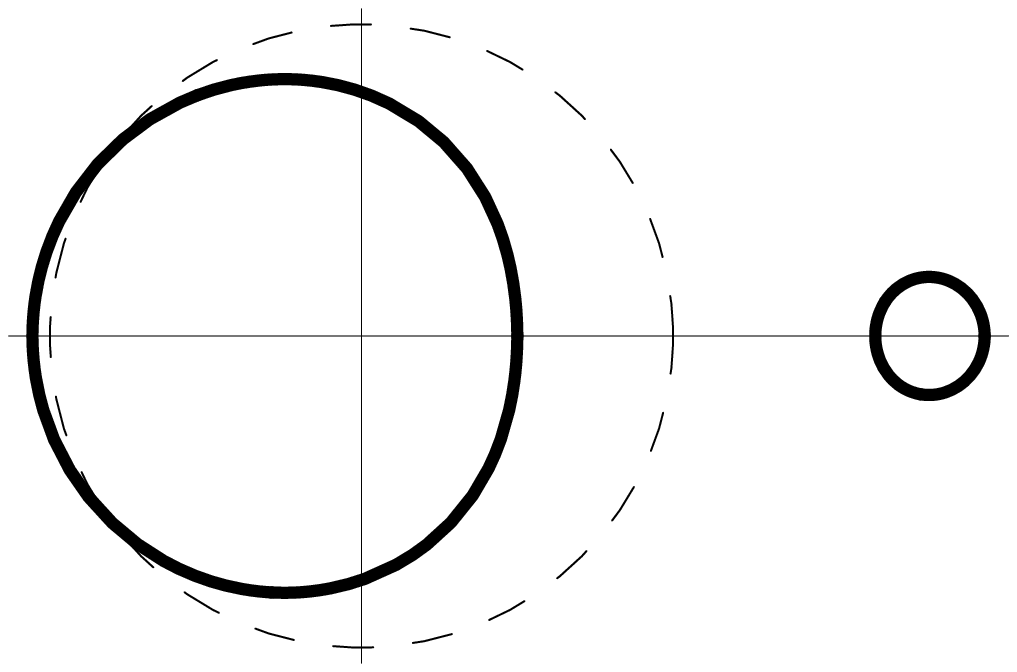 , width=3.5cm}
{\normalsize Fig. 2(a)}
\end{minipage} \hfill
\begin{minipage}[b]{.46\linewidth}
\centering\epsfig{figure=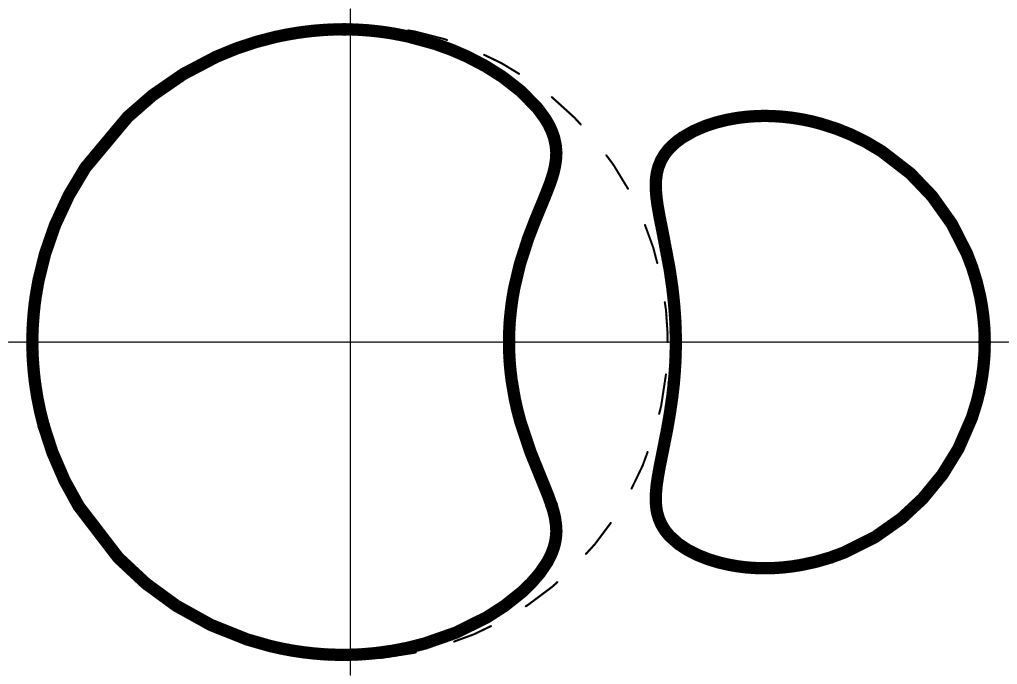 , width=3.5cm}
{\normalsize Fig. 2(b)}
\end{minipage}

$\ $\\
$\ $

\begin{minipage}[b]{.46\linewidth}
\centering\epsfig{figure=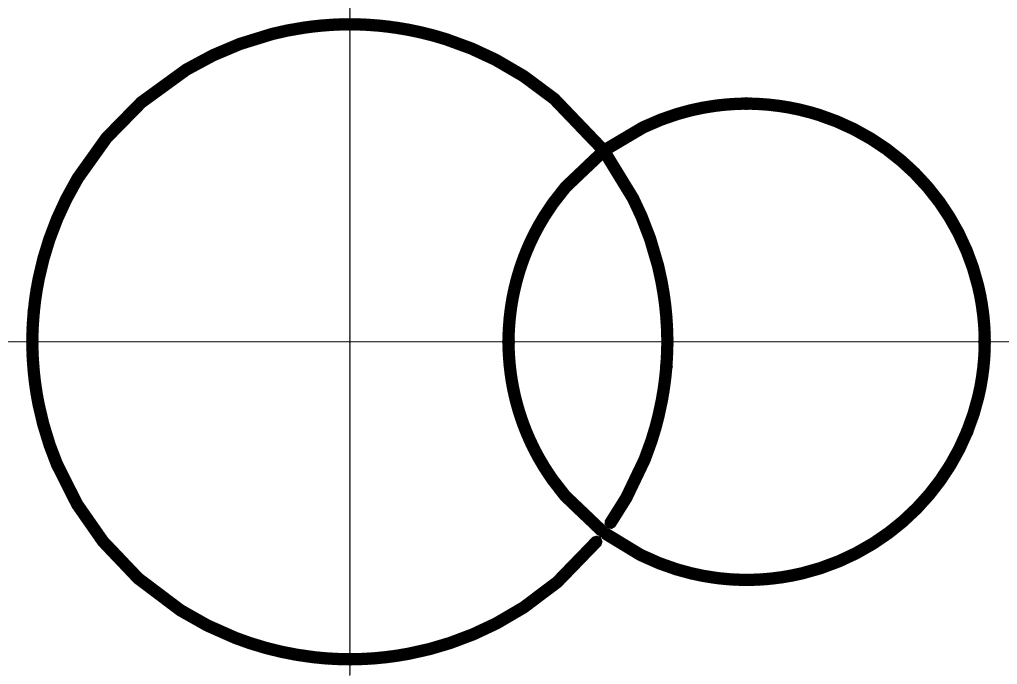 , width=3.5cm}
{\normalsize Fig. 2(c)}
\end{minipage} \hfill
\begin{minipage}[b]{.46\linewidth}
\centering\epsfig{figure=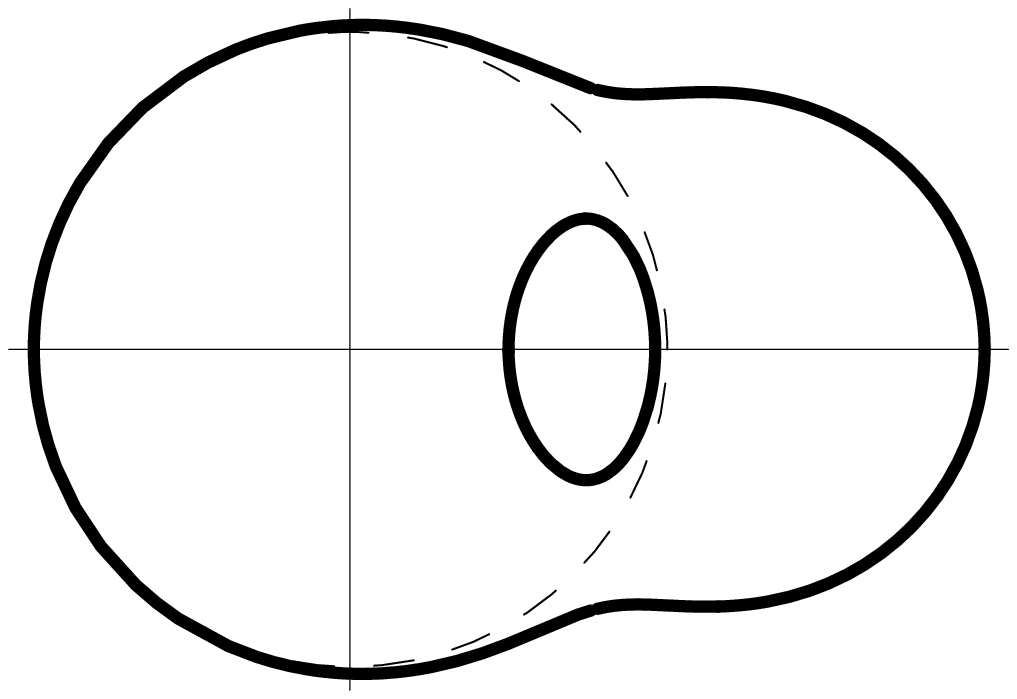 , width=3.5cm}
{\normalsize Fig. 2(d)}
\end{minipage}
$\ $\\

{\large \bf Figure 2 :}\\
Except for 2(c), no collisions occur, and each vortex returns
to its initial position after one period. The degenerate case 2(c)
is the same as 1(c) --- one great circle and another circle --- in
a different orientation.

$\ $\\

\begin{center}
\epsfig{figure=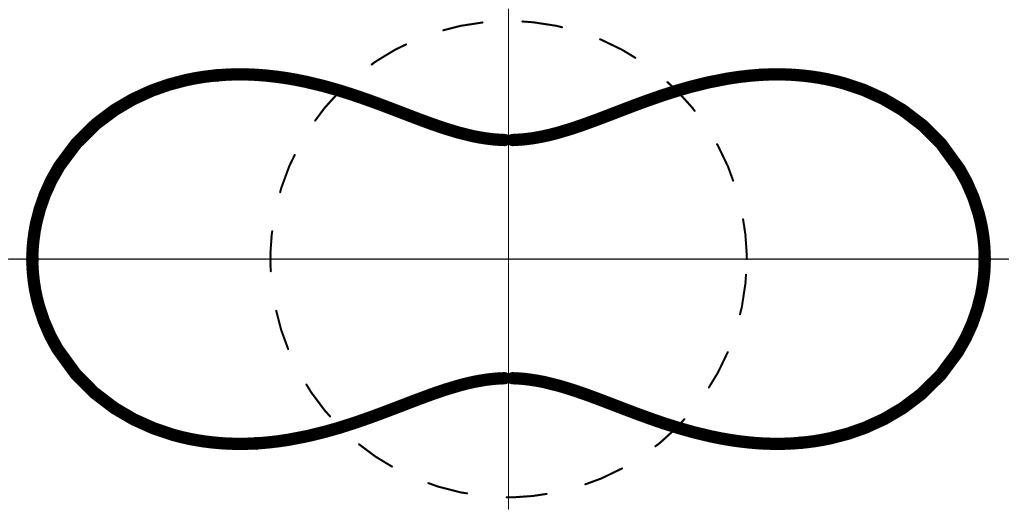 , width=5cm}
{\normalsize Fig. 3}
\end{center}

$\ $

{\large {\bf Figure 3 ({\sl with a=2}) :}}\\ 
No collision takes place and the vortices exchange positions after one 
period. The coordinate $c^1(t)$ of expressions $(14)$ and $(18)$ is always
zero. The coordinate $c^2(t)$ is of the form\\ $
-a^2 + 2a(1+a^4)/[2a^3 + i \omega\, \cot (\omega t)]$ .
Taking one of the roots $z(t)\,=\,x(t)\,+\,i\,y(t)$ of $(18)$ and
eliminating $t$ from the system $x(t),y(t)\,$, one obtains that the
trajectory on the plane has the simple equation $(x^2 + y^2)^2 +
(1/a^2 - a^2)(x^2 - y^2) - 1 \,=\,0$. These are special cases of
Cassini's ovals and, when projected back to the sphere, look
like the joint of a tennis ball.

\section{Coincident particles and collisions}

  In this Section we start by finding an algebraic condition which
  determines when a point in the moduli space ${\mathcal M}_N \cong
  \Bbb{CP}^N$ corresponds to a vortex configuration where at least
  two of the vortices coincide. We then use this condition to show that, for a
  system of $N$ vortices starting in different positions with
  arbitrary initial velocities, there are at most $2N-2$ collisions
  during one period of the motion.\\
$\ $

Using diagram $(9)$ and the definition $(8)$ of the bijection
$\alpha$, it is not difficult to recognize that a point
$[w_0,\ldots,w_N] \in \Bbb{CP}^N$ corresponds by $\tilde{\alpha}^{-1}$
to the equivalence class in ${\mathcal M_N}$ of a solution $\phi$ represented by 
\[
\phi_1(z)\, =\, A\,(1+|z|^2)^{-N/2}\,\sum_{k=0}^N
\frac{w_k}{(k!(N-k)!)^{1/2}}\, z^{N-k}\ ,
\]
where $A$ is a non-zero normalization factor. Thus, asserting that
$[w_0,\,\ldots ,\, w_N]$ corresponds to a solution with at least
two coincident vortices is equivalent to saying that one of the
following conditions holds :
\begin{align*}
&\bullet \ P(z) \,:=\, \sum_{k=0}^N \frac{w_k}{(k!(N- k)!)^{1/2}}\,
z^{N-k}\ \ \text{has a root of multiplicity at least two}\, ; \\
&\bullet \ w_0\, =\, w_1\, =\ 0\  ,\text{ which corresponds to a double zero of
$\phi$ at}\ [0,1] \in \Bbb{CP}^1\ .
\end{align*}
We now use the following result, whose proof is at the end of this
section :\\
$\ $\\
 {\bf Lemma:} {\it 
For any $n \in \Bbb{N}$, there is a homogeneous polynomial $S$ in $n+1$
variables of degree $2n-2$, such that $S(a_0,
\ldots , a_n)=0 $ if and only if $\,\sum_{k=0}^{n} a_k z^{n-k}$ has a
multiple root or $\,a_0 = a_1 = 0\, .$  } \\
$\ $\\
An explicit formula for $S$ is given in the proof and we note that,
up to a sign, $S$ coincides with the discriminant of
$\,\sum_k a_k z^{N-k}$ whenever $a_0 \neq 0 $ .

Using this lemma, it is clear that the points $[w_0,\ldots,w_N]
\in \Bbb{CP}^N$ which correspond to at least two coincident vortices,
are exactly those of the algebraic hypersurface $H$ of degree $2N-2$
in $\Bbb{CP}^N$ determined by the equation 
\[  
\tilde{S}(\ldots , w_k, \ldots )\, :=\,
 S(\,\ldots\,,\, \frac{w_k}{(k!(N-k)!)^{1/2}}\,,\, \ldots\,)\ =\ 0\ .
\]

 As we have seen in Section $5$, any motion of $N$ vortices
in $S^2$ corresponds to a Fubini-Study geodesic in $\Bbb{CP}^N$, and
these are all of the form $t \mapsto \pi(\sin (\omega t)y + \cos
(\omega t)x)$, with $x,y \in \Bbb{C}^{N+1}$ orthonormal and $\pi$
being the  projection from $\Bbb{C}^{N+1}$ to $\Bbb{CP}^N$. By the
discussion above, it is also clear that this motion will have a
collision of two or more vortices iff the corresponding geodesic
intersects $H$. But since this geodesic lies on the projective line
$L\,= \pi (\text{span}_{\Bbb{C}} \{x,y\}) \subset \Bbb{CP}^N\,,$ and
does not intersect itself during one period, we conclude that the
number of collisions is not bigger than the cardinality of $L\cap H$.

It is, however, a standard result in algebraic geometry that either
$L\subset H\,$ or $\,$\# $\!(L\cap H)\, \leq \text{deg}\,H\,=\,2N-2\ .$  
In fact, denoting $x=(x_0,\,\ldots ,x_N)$ and $y=(y_0,\,\ldots ,y_N)$ in
$\Bbb{C}^{N+1}$, it follows that a point $\pi (ux+vy)$ in $L$, with
$(u,v)\in \Bbb{C}^2\!\setminus \!\!\{0\}$, belongs to $H$ iff
\[  Q(u,v)\ :=\ \tilde{S}(ux_0+vy_0, \ldots , ux_N +vy_N)\ =\ 0\ . \]
But since $\tilde{S}$ is homogeneous of degree $2N-2$, so is $Q$, and 
therefore there is a factorization (see \cite[p. 31]{Kir} )
\[
Q(u,v)\ =\ \prod_{i=1}^{2N-2} (\alpha_i u + \beta_i v)\ \ ,\qquad \ \
\ \text{for some }(\alpha_i,\beta_i)\in\Bbb{C}^2\ .
\]
If $Q$ is identically zero we have $L\subset H$. If $Q$ is not
identically zero then the roots of $Q$ are $(\beta _i,-\alpha _i)
\neq 0 \ \ \forall i\,$, and it is apparent that $L\cap H$
consists of the points $\pi (\beta_i x -\alpha_i y)$, which are at most $2N-2$.

We finally conclude that, either the vortices have a motion with at
least two of them coincident for all time, which corresponds to
$L\subset H$, or there are at most $2N-2$ collisions in one period.
\\

$\ $\\
{\bf Proof of the lemma :}\\
This lemma is a slight generalization of well-known algebraic results,
as stated for example in \cite[p. 168]{Gri} or \cite[p. 178]{Co}.
  Consider $P(z)=\sum_{k=0}^n a_k\,z^{n-k}$ and its derivative
  $P'(z)=\sum_{k=1}^n (n-k+1)a_{k-1}z^{n-k}$, and form the usual resultant
\[
{\mathcal R}_{P,P'}(a_0,\ldots ,a_n) := 
\begin{vmatrix}
a_0 & \       & \cdots & \      & \ \ a_n \\
\   & \ddots  & \      &  \     & \           &\ddots   \\
\   & \       & a_0    & \      & \cdots      &\       & a_n  \\
na_0& \       & \cdots & \      & a_{n-1}  \\
\   & \ddots  & \      & \      & \           & \ddots   \\
\   & \       & \ \ \ \ na_0   & \      & \cdots      & \      & a_{n-1}
\end{vmatrix}
\] 
where there are $n-1$ rows with the coefficients of $P$ and $n$ rows
with the coefficients of $P'$, so that the matrix is $(2n-1)\times
(2n-1)$. Applying the usual expansion to the first column of this
determinant we get 
\[ {\mathcal R}_{P,P'}(a_0,\ldots ,a_n) = a_0\, S(a_0, \dots , a_n) \]
with
\begin{multline*}
S(a_0,\ldots , a_n) \ = \\
=\ 
\begin{vmatrix}
a_0     & \cdots   & \      & a_n    \\
\       & \ddots   & \      & \       &\ddots  \\
\       &   \      & a_0    & \cdots  & \     & a_n  \\ 
(n-1)a_1& \cdots   & a_{n-1} \\
na_0    & \cdots   & \      & a_{n-1}  \\
\       & \ddots   & \      & \       & \ddots  \\
\       & \        & na_0   & \cdots  & \     & a_{n-1}
\end{vmatrix}
\ + \ (-1)^n\, n\,
\begin{vmatrix}
a_1  & \cdots  & a_n  \\
a_0  & \cdots  & \    & a_n   \\
\    & \ddots  & \    & \      & \ddots   \\
\    & \       & a_0  & \cdots & \      & a_n  \\
na_0 & \cdots  & \    & a_{n-1}   \\
\    & \ddots  & \    & \      & \ddots  \\
\    & \       & na_0 & \cdots & \      & a_{n-1}  
\end{vmatrix}\ \ .
\end{multline*}
Expanding again the first columns of the
determinants in $S(0, a_1,\ldots ,a_n)$ we get
\[
S(0,a_1,\ldots , a_n)\  =\  (-1)^n (n-1)\,a_1\, {\mathcal R}_{Q,Q'} + (-1)^n n
\,a_1\, {\mathcal R}_{Q,Q'}\ =\ (-1)^n (2n-1)\, a_1\, {\mathcal R}_{Q,Q'}
\]
where $Q(z)= \sum_{k=1}^n a_k \,z^{n-k}$.

Now, if $a_0 \neq 0\,$, by standard results in basic algebra (see
\cite{Gri,Co}),
\[  {\mathcal R}_{P,P'} =  (-1)^{n(n-1)/2} a_0\, D(P)\ , \]
where $D(P)$ is the discriminant of $P$, and therefore 
\[
S(a_0,\ldots ,a_n)=0\ \ \Leftrightarrow\ \ D(P)=0\ \ \Leftrightarrow\
\ P\text{ has a multiple root}\ .
\]
If $a_0=0$ then $P(z)=Q(z)$ and 
\[
S(a_0,\ldots ,a_n) = S(0,a_1,\ldots ,a_n)=0\ \ \Leftrightarrow\ \ a_1=0
\text{ or } {\mathcal R}_{Q,Q'}=0 \ .
\]
But when $a_1\neq 0$, by the same algebraic results, ${\mathcal
R}_{Q,Q'}=0\, \Leftrightarrow\, Q=P\ \text{has a multiple root}$.

We finally conclude that 
\[
S(a_0,\ldots ,a_n)=0\ \ \Leftrightarrow \ \ P \text{ has a multiple root or
} a_0=a_1=0 \ .
\]

\vspace{1.5cm}
\vskip 20pt
{\centerline{\bf{Acknowledgements}}}
\vspace{.25cm}
\noindent
NSM thanks Michael Singer for discussions about the Fubini-Study
metric, and also thanks the Pure Mathematics Department, University of
Adelaide, for hospitality at the time this work was initiated. JMB is
supported by the \lq{\sl Funda\c{c}\~ao para a Ci\^encia e
Tecnologia}\rq, Portugal,
through the research grant SFRH/BD/4828/2001.

\newpage
\noindent
{\Large {\bf Appendix}}\\
$\ $\\
$\ $

In this appendix we compute integrals of the form
\[ \int_{\Bbb{C}} \phi\, \bar{\psi}\, \frac{2iR^2}{(1+|z|^2)^2}\, dz\wedge d\bar{z} \ =\ 
   \int_{\Bbb{R}^2} \phi\, \bar{\psi}\, \frac{4R^2}{(1+x^2+y^2)^2}\, dx\, dy     \]
where $\phi = (1+|z|^2)^{-N/2}(a_0 z^N + \cdots + a_N)$ and $\psi = (1+|z|^2)^{-N/2}
(b_0 z^N+ \cdots + b_N)$ .

Write 
\[ \phi\,\bar{\psi}\, =\, \sum_{k,j=0} ^N a_{N-j}\, \bar{b}_{N-k}\,f_{kj}(z)\ \ ,\ \ \ \ 
\ \ \ \ \ \ \text{with}\ \ \ \ \ \ f_{kj}(z) = \bar{z}^k z^j (1+|z|^2)^{-N} \ . \]
Using polar coordinates and integration by parts,
\begin{align*}
I(k,j,N) :& =\ \int_{\Bbb{R}^2} \frac{f_{kj}(z)}{(1+|z|^2)^2}\ =\ 
\int_0 ^{2\pi} e^{i(j-k)\theta}\, d\theta\ \ \int_0 ^{\infty} \frac{r^{k+j}}
{(1+r^2)^{N+2}}\,r\, dr \ = \\
& =\  2\pi\,\delta_{jk} \int_0 ^{\infty} \frac{r^{2k+1}}
{(1+r^2)^{N+2}}\, dr \,=\, \delta_{jk} \frac{-\pi}{N+1} \int_0 ^{\infty} r^{2k} \frac{d}{dr}
 \left( \frac{1}{(1+r^2)^{N+1}} \right) \, dr\ = \\
& =\ \delta_{jk}\,\frac{2k\pi}{N+1} \int_0 ^{\infty} \frac{r^{2k-1}}{(1+r^2)^{N+1}} \, dr\ 
 =\ \delta_{jk}\, \frac{k}{N+1}\, I(k-1, k-1, N-1)
\end{align*}
where the vanishing of the boundary terms in the integration by parts is valid for 
$N\geq k \geq 1$. Since 
\[ 
I(0,0,N-k)\,=\, 2\pi \int_0 ^{\infty} \frac{r}{(1+r^2)^{N-k+2}} \,dr\,=\, \frac{-\pi}{N-k+1}
\ \bigg[\frac{1}{(1+r^2)^{N-k+1}}\bigg]_0 ^{\infty} =\, \frac{\pi}{N-k+1}  
\]
we have
\[  I(k,j,N)\,=\, \delta_{jk}\,\frac{k(k-1)\cdots 1}{(N+1)\cdots (N+2-k)}\ I(0,0,N-k) \,=\, 
     \frac{k!(N-k)!}{(N+1)!}\, \pi\, \delta_{jk}\ \ .
\]
The final result is therefore
\[
\int_{\Bbb{C}} \phi\, \bar{\psi}\, \frac{2iR^2}{(1+|z|^2)^2} \, dz\wedge d\bar{z} \ =\ 
4R^2\,\sum_{k=0} ^N \,I(k,k,N)\, a_{N-k}\, \bar{b}_{N-k} \,=\, 4\pi R^2 \,\sum_{k=0} ^N 
\, \frac{k!(N-k)!}{(N+1)!} \,a_k \,\bar{b}_k \ \ .
\]

\newpage

\end{document}